\documentclass[11pt]{article}
\pdfoutput=1
\usepackage{jheparxiv}

\usepackage[dvipsnames]{xcolor}

\usepackage{bm}
\usepackage{amsmath}
\usepackage{amssymb}
\usepackage{graphicx}
\usepackage{slashed}
\usepackage{amsfonts}
\usepackage{comment}

\usepackage{tikz-feynman}

\usepackage{hyperref}
\hypersetup{
colorlinks=true,         
linkcolor=blue,          
citecolor=red,        
urlcolor=blue            
}

\newcommand{\be}{\begin{equation}}
\newcommand{\ee}{\end{equation}}
\newcommand{\bi}{\begin{itemize}}
\newcommand{\ei}{\end{itemize}}
\newcommand{\bea}{\begin{eqnarray}}
\newcommand{\eea}{\end{eqnarray}}
\newcommand{\ud}{\mathrm{d}}

\newcommand{\sfT}{{\mathsf T}}          
\newcommand{\sfd}{{d}}          

\newcommand{\pd}{\partial}
\newcommand{\order}[1]{\mathcal{O}\left(#1 \right)}

\newcommand{\deltahat}{\hat{\delta}}

\newcommand{\h}{\mathfrak{H}}

\begin{document}

\title{Toward double copy on arbitrary backgrounds}

\author{Anton Ilderton}
\emailAdd{anton.ilderton@ed.ac.uk}

\author{and William Lindved}
\emailAdd{william.lindved@ed.ac.uk}

\affiliation{Higgs Centre, University of Edinburgh, EH9 3FD, Scotland, UK}

\abstract{
Double copy relates scattering amplitudes in a web of gravitational and gauge theories. Although it has seen great success when applied to amplitudes in vacuum, far less is known about double copy in arbitrary gravitational and gauge backgrounds. Focussing on the simplest pair production amplitudes of scalar QCD in a background gauge field, we construct, at next-to-leading order in perturbation theory, a double copy map to particle production in general metrics (and associated axio-dilatons) constructed from the gauge background. We connect our results to convolutional and classical double copy and, turning to examples, identify a class of gauge fields which generate FRW spacetimes via double copy. For this case we are able to conjecture the all-orders form of the double copy map.
}

\maketitle

\section{Introduction}
Double copy relates scattering in gauge theory to scattering in gravity, by replacing the colour structure of amplitudes with a copy of the kinematic structure~\cite{Kawai:1985xq,Bern:2008qj,Bern:2010ue,Bern:2010yg,Cachazo:2013gna,Cachazo:2013hca}. As well as offering a practical method to streamline gravitational scattering calculations by mapping them to simpler gauge theory calculations, it suggests a deep connection between gauge and gravitational theories. For reviews see~\cite{Bern:2019prr,Bern:2022wqg,Adamo:2022dcm,White:2024pve}.

Double copy has been extremely successful when applied to amplitudes in vacuum. If it is a fundamental relationship between gauge theory and gravity then double copy should presumably hold in some form on any background, and not just for perturbative scattering on trivial backgrounds, i.e.~vacuum. There are many physical situations, in both gravity and gauge theory, which motivate the study of background fields, the presence of which often leads to explicitly non-linear and non-perturbative phenomena. Examples include gravitational memory~\cite{Christodoulou:1991cr}, QCD in strong magnetic fields~\cite{Miransky:2015ava,Hattori:2016emy}, magnetar environments~\cite{Turolla:2015mwa,Kaspi:2017fwg,Kim:2021kif}, lab-astrophysics~\cite{Marcowith,Kim:2019joy} and intense laser physics~\cite{Fedotov:2022ely}.

In asking how double copy looks on arbitrary backgrounds there are, very broadly, two directions to consider. First, one can ask whether gauge theory amplitudes in a given spacetime map to gravity amplitudes in the same spacetime; this is (implicitly) what is done, and works, for perturbation theory in vacuum, where the background is Minkowski. Here quite some progress has been made in the case of AdS space and related CFT correlators~\cite{Farrow:2018yni,Lipstein:2019mpu,Alday:2021odx,Diwakar:2021juk,Jain:2021qcl,Zhou:2021gnu,Mei:2023jkb,Lipstein:2023pih}.

Second, one can ask whether scattering on a given gauge background maps to scattering in some double-copy-related spacetime. Again progress can be made by turning to specific examples; a double copy map has been found which relates scattering on gauge and gravitational plane waves~\cite{Adamo:2017nia,Adamo:2017sze,Adamo:2020qru}, which holds \emph{non}-perturbatively in the coupling to the background. The study of conserved quantities in highly symmetric backgrounds has also revealed double copy maps for geodesics and related observables between e.g.~$\sqrt{\text{Schw}}$ and Schwarzschild, $\sqrt{\text{Kerr}}$ and Kerr~\cite{Gonzo:2021drq}, and between electromagnetic vortices and gravitational waves~\cite{Ilderton:2018lsf}. For \emph{classical} double copy in curved spacetime see e.g.~\cite{Bahjat-Abbas:2017htu,Carrillo-Gonzalez:2017iyj,Han:2022mze,Prabhu:2020avf,Alkac:2021bav}.

{We pursue the second of these options in this paper. Our approach is simple but direct;} focussing on fundamental matter coupled to Yang-Mills (scalar QCD), and its double copy~\cite{Johansson:2014zca,Johansson:2015oia,Plefka:2019wyg}, {we will calculate the simplest gauge theory scattering amplitudes of the dynamical scalar field} in the presence of {asymptotically flat, but otherwise \emph{arbitrary}, gauge backgrounds $A^\text{b.g.}$. We will then construct a double copy map whose output is the corresponding scattering amplitude for dynamical scalar fields on a gravitational background $g\sim \eta +A^\text{b.g.}A^\text{b.g.}$, plus associated axiodilaton fields.} {The considered scattering processes are two-point amplitudes at tree level -- unlike} in vacuum, such amplitudes are highly non-trivial on backgrounds, as they encode geodesic motion of test particles, associated memory effects, and also particle production. Further, these amplitudes can be arbitrarily complex even at tree level because of the in-principle arbitrary \emph{functional} degrees of freedom in the background (i.e.~tree amplitudes will not typically be rational functions of kinematic data).

While one aim of this investigation is to, eventually, identify double-copy structures in the \emph{non}-perturbative parts of scattering amplitudes on backgrounds, we will begin in this paper by treating all interactions with the background perturbatively. Calculating order by order in perturbation theory we will find non-trivial relations even at leading order in the coupling, since three-point tree amplitudes are kinematically allowed in a background (and not just for zero momentum bosons).

This paper is organised as follows. In Sec.~\ref{sec:leading} we introduce the theories of interest, namely (fundamental) scalars coupled to Yang-Mills, and to axio-dilaton gravity. We calculate the leading order contribution to particle production in arbitrary backgrounds, and identify a double copy map between the gauge and gravitational theories, with the respective backgrounds related by the `{convolutional double copy'~\cite{Anastasiou:2014qba,Anastasiou:2018rdx,Luna:2020adi,Borsten:2019prq,Borsten:2020xbt,Liang:2023zxo}. In} Sec.~\ref{sec:NLO} we extend this calculation, and the double copy map, to next-to-leading order (NLO), i.e.~beyond the linearised theory. Here the nonlinearity of Einstein's equations makes itself known, imposing constraints on the relation between the gauge and gravitational backgrounds if the double copy is to hold order by order in perturbation theory. 

In Sec.~\ref{sec:examples} we conjecture an all-orders form for the double copy. We then analyse the dilaton contribution to our 2-point amplitudes, and show non-perturbatively that it is equivalent to a non-conformal matter coupling, which greatly simplifies computations. Turning to examples, we focus on pair creation in FRW spacetimes,  and in time-dependent (chromo) electric fields. The simplifications offered by going to specific choices of background allow us to extend our calculations to third perturbative order. This reveals some potential subtleties which may arise beyond NLO. It also allows us to conjecture an explicit all-orders form for the double copy map to FRW spacetimes, and to revisit and reinterpret the constraints imposed for general backgrounds at NLO. We conclude in Sec.~\ref{sec:conclusions}.

\subsection*{Notation and conventions}
$\hbar=1$ throughout. Mostly minus metric. Factors of $2\pi$ in delta functions and integration measures are absorbed into hats, so $\hat \delta = (2\pi) \delta$ and $\hat \ud = \ud/(2\pi)$.
\section{Leading order}\label{sec:leading}

\subsection{Gauge theory}
%
%
We consider a massive scalar field $\varphi(x)$, in the fundamental {representation}, minimally coupled to a YM field $A_\mu(x) \equiv A_\mu^a(x) \mathsf{T}^a$ with $\mathsf{T}^a$ the group generators. The Lagrangian is simply
\be
    \mathcal{L} =  -\frac14 F_{\mu\nu}^a F^{a\mu\nu} + (D_\mu\varphi)^\dagger D^\mu \varphi -m^2 \varphi^\dagger \varphi \,,
\ee
in which the covariant derivative is $D_\mu = \partial_\mu - i g A_\mu$. We introduce a background by separating $A_\mu \to A^\text{b.g.}_\mu + A_\mu$, in which $A_\mu$ is now to be quantised, along with $\varphi$, on the background $A_\mu^\text{b.g.}$. This background may be sourced or not, and we make no assumptions on its spacetime dependence, {except for the assumption that the potential vanishes asymptotically, so that we have a well-defined notion of asymptotic states}.

{We have in mind physical situations where the background is strong, i.e.~the coupling of the matter field to the background $\propto gA^\text{b.g.}/m\gg~g$. Thus we work with a truncation of the theory where we neglect fluctuations of the gauge field and consider a scalar in the background with Lagrangian}
\be\label{gauge-reduced-action}
    \mathcal{L} \to 
    (D_\mu\varphi)^\dagger D^\mu \varphi -m^2 \varphi^\dagger \varphi \,,
\ee
where now $D_\mu = \partial_\mu - i g A^\text{b.g.}_\mu$. We drop `b.g.' from the gauge field from here on. 

As the only dynamical field in the theory is now $\varphi$, there are only two interesting amplitudes; the $1\to 1$ amplitude for a massive particle scattering on the background, and the $0\to 2$ pair production amplitude. Formally these are related by a change in sign for one of the particle momenta, but they obviously differ physically; the $0\to 2$ amplitude is purely quantum, for example, and for the sake of concreteness we focus on this throughout.

As discussed in the introduction, the eventual aim is to make non-perturbative statements about double copy in backgrounds, but we will begin here by working perturbatively in the coupling to the background.
As such, the Feynman rules of our truncated theory (\ref{gauge-reduced-action}) are given in Fig.~\ref{FIG:FEYN-YM}; note that at every vertex we conserve momentum and integrate ${\hat \ud}^4k$ for each background field momentum. The leading order contribution to the pair production amplitude, call it $\mathcal{A}_{(1)}$, is simply the three-point diagram with the massive legs taken on-shell. One immediately finds
\be\label{YM1}
     {i \mathcal{A}_{(1)} = -(2g)}\int\!\hat{\ud}^4 k\, \hat{\delta}^4(p + q - k) A^{a \mu}(k) \sfT^a_{ij}  {\big(p    - \tfrac12 k\big)}_\mu \;,
\ee
in which $p_\mu$ and $q_\mu$ are the momenta of the produced pair, while $\sfT^a_{ij} \equiv c_i \sfT^a c_j$ with $c$ the colour vectors of the pair. The amplitude is easy to understand; it is a three-point amplitude in vacuum, with an off-shell gluon, integrated against the background field profile. This remaining integral is trivially evaluated, setting the argument of the gauge field (the momentum influx from the field) equal to the produced particle momentum, as expected. We leave the integral unevaluated as this will later help expose the double copy structure, just as in vacuum. 

\begin{figure}[t!]
\begin{equation*}
	\begin{tikzpicture}[baseline=(o.base)]
		\begin{feynman}[small]
			\vertex (o) at (0,0);
			\vertex[above right=of o] (s1);
			\vertex[below right=of o] (s2);
			\vertex[left=of o] (b){$\times$};			
			\diagram* {
				(o) -- [fermion, momentum = $p$] (s1),
				(o) -- [anti fermion, momentum' = $q$] (s2),
				(b) -- [gluon] (o),
			};
		\end{feynman}
	\end{tikzpicture}
	= i g \sfT^a  A^a(k) \cdot (p - q) \qquad \quad
	\begin{tikzpicture}[baseline=(o.base)]
		\begin{feynman}[small]
			\vertex (o) at (0,0);
			\vertex[above right=of o] (s1);
			\vertex[below right=of o] (s2);
			\vertex[above left=of o] (b1){$\times$};
			\vertex[below left=of o] (b2){$\times$};			
			\diagram* {
				(o) -- [fermion, momentum = $p$] (s1),
				(o) -- [anti fermion, momentum' = $q$] (s2),
				(b1) -- [gluon] (o),
				(b2) -- [gluon] (o),
			};
		\end{feynman}
	\end{tikzpicture} 
	= i g^2 \sfT^a \sfT^b A^a(k_1) \cdot A^b(k_2),
	%
\end{equation*}
\caption{\label{FIG:FEYN-YM} The Feynman rules for a massive (fundamental) scalar particle coupled to the background YM field $A_\mu$, indicated by a cross. Massive particle momenta are outgoing.
} 
\end{figure}

No gauge has been assumed. It can be checked without difficulty that (\ref{YM1}) is invariant, on the support of the momentum-conserving delta function, under linearised gauge transformations or, equivalently, invariant up to terms of order $g^2$ under the full gauge transformation, here parameterised by $\omega(x)$,
\be\label{gauge-x-form-full}
    A_\mu(x) \to A_\mu(x) + \partial_\mu \omega(x) -i g [A_\mu(x),\omega(x)] \;.
\ee
%
\subsection{Gravity}
%
Turning to gravity, we begin with the conjectured all-order action for the double copy of a (single flavour of) fundamental scalar coupled to Yang-Mills, as given in~\cite{Plefka:2019wyg}; this is 
\begin{equation}
\label{DC-L-1}
	S = \int\!\ud^4 x\, \sqrt{-g} \bigg(
 {\frac{1}{\kappa^2}\bigg[-2R +\frac{g^{\mu\nu} \pd_\mu{\bar Z} \pd_\nu Z}{\big(1 - \tfrac14 {\bar Z}Z\big)^2}\bigg]}
 + \frac{g^{\mu\nu} \pd_\mu \varphi^\dagger \pd_\nu \varphi - m^2
 e^{-\phi}
 \varphi^\dagger \varphi}{\big( 1 - \frac{\kappa^2}{32} \varphi^\dagger \varphi   \big)^2}\bigg)\;,
\end{equation}
in which the matter field $\varphi$ is a complex scalar, while $Z$ encodes the dilaton $\phi$ and axion $\chi$ as
\be
    Z \equiv 2\frac{\chi+i (e^{-\phi}-1)}{\chi+i (e^{-\phi}+1)} \;.
\ee
We have chosen alternative scalings of the latter fields, compared to the literature, which will make clearer the separation into background and dynamical pieces. 

{The action  (\ref{DC-L-1}) was derived by perturbatively matching gauge theory and gravity amplitudes, and then conjecturing the all-orders form shown based on observed patterns. We note the striking similarity of the contact terms in the matter sector to the structures in the axio-dilaton gravity} sector~\cite{Plefka:2019wyg}. We will later see similar structures appearing through double copy in the purely gravitational sector.

We perform a field redefinition $\bar g_{\mu\nu} = e^{- \phi} g_{\mu\nu}$ (hence $\bar g^{\mu\nu}  = e^{\phi} g^{\mu\nu}$ and $\bar g = g e^{-4 \phi}$), such that the dilaton couples to all matter terms in the same way, which allows for a particularly clear double copy prescription. Inserting this into (\ref{DC-L-1}) and transforming the Ricci scalar accordingly~\cite[App.~G]{Carroll:2004st}, the action (\ref{DC-L-1}) becomes, immediately dropping the bar on the new metric for clarity and writing out the dilaton and axion terms explicitly~\cite{Johansson:2019dnu},
\begin{equation}
\label{DC-L-2}
	S = \int\!\ud^4 x\, \sqrt{-g} e^{\phi} \bigg[
 \frac{-2R -2 g^{\mu\nu} \pd_\mu \phi \pd_\nu \phi +   e^{\phi} g^{\mu\nu} \pd_\mu \chi \pd_\nu \chi }{\kappa^2}
 + \frac{g^{\mu\nu} \pd_\mu \varphi^\dagger \pd_\nu \varphi - m^2 \varphi^\dagger \varphi}{\big(1 - \tfrac{\kappa^2}{32}\varphi^\dagger \varphi\big)^2} 
 \bigg].
\end{equation}
To make the equivalent truncation as in gauge theory we separate the metric, dilaton and axion into backgrounds and dynamical fluctuations, subsequently dropping the latter.
We  {then} expand the background around flat space, writing
\be
    g^{\mu\nu} = \eta^{\mu\nu} - \kappa h^{\mu\nu} \;, \qquad \phi = \kappa \phi_{(1)} \;, 
    \qquad \chi = \kappa \chi_{(1)} \;.
\ee
Since $\kappa$ is negligible compared to the stronger coupling to the background,  $\kappa h$, we can also drop the contact terms.
The surviving part of the action is then (\ref{DC-L-2}) without the contact terms and where all axio-dilaton gravity fields are interpreted as backgrounds. 
Only one term in the action contains dynamical fields, namely
\begin{equation}
\label{quad-action}
 \int\!\ud^4 x\,  \sqrt{-g}\, {e^{\phi}} \big( g^{\mu\nu} \pd_\mu \varphi^\dagger \pd_\nu \varphi - m^2 \varphi^\dagger \varphi \big)\;,
\end{equation}
and we note that the axion does not couple directly to the matter field.

\begin{figure}[t!]
\begin{equation*}
\begin{split}
	\begin{tikzpicture}[baseline=(o.base)]
		\begin{feynman}[small]
			\vertex (o) at (0,0);
			\vertex[above right=of o] (s1);
			\vertex[below right=of o] (s2);
			\vertex[left=of o] (b){$\times$};
			\diagram* {
				(o) -- [fermion, momentum = $p$] (s1),
				(o) -- [anti fermion, momentum' = $q$] (s2),
				(b) -- [photon] (o),
			};
		\end{feynman}
	\end{tikzpicture} 
	&
 = i  \kappa h^{\mu\nu}(k) \left[ p_\mu q_\nu   -  \frac12 \left(p \cdot q + m^2  \right)\eta_{\mu\nu}     \right]
 \\
	\begin{tikzpicture}[baseline=(o.base)]
		\begin{feynman}[small]
			\vertex (o) at (0,0);
			\vertex[above right=of o] (s1);
			\vertex[below right=of o] (s2);
			\vertex[left=of o] (b){$\times$};
			\diagram* {
				(o) -- [fermion, momentum = $p$] (s1),
				(o) -- [anti fermion, momentum' = $q$] (s2),
				(b) -- [scalar] (o),
			};
		\end{feynman}
	\end{tikzpicture}
	&= - i \kappa \phi_{(1)}(k) \left( p \cdot q + m^2  \right), 
\end{split} 	
\end{equation*}
\caption{\label{FIG:FEYN-GR1} The Feynman rules for a massive (complex) scalar particle coupled to a background metric perturbation $h^{\mu\nu}$, and dilaton $\phi$, in the linearised approximation.} 
\end{figure}

We assume that the metric, like the gauge field, is asymptotically flat~\cite{Adamo:2021dfg}, and moreover that $h\to0$ asymptotically -- relaxing this is not difficult, as we comment on later, but it is a natural starting point when working perturbatively. The Feynman rules are given in Fig.~\ref{FIG:FEYN-GR1} and the leading order amplitude is
\be\label{GR1}
i\mathcal{M}_{(1)} = -\kappa \int\!\hat{\ud}^4 k\, \hat{\delta}^4(p + q - k) \bigg[ h^{\mu\nu}(k)\Big( p_\mu q_\nu - \frac12 (p\cdot q +m^2)\eta_{\mu\nu}\Big) - \phi_{(1)}(k)(p\cdot q +m^2) \bigg]\;.
\ee
The metric contribution can be checked against the literature; the pair creation amplitude and probability were discussed extensively in\footnote{The non-minimal coupling included in~\cite{Frieman:1985fr} seems to change sign between the Lagrangian and the Feynman rules. We mention this only because a non-minimal coupling will appear below, and the sign is important.}~\cite{Frieman:1985fr}. Again, no gauge has been chosen for the metric. Invariance of (\ref{GR1}) under linearised diffeomorphisms is easily verified. 
\subsection{Double copy}
%
In this section we identify a double copy relation between the gauge and gravitational amplitudes (\ref{YM1}) and (\ref{GR1}). There are essentially two questions to ask; first, how can the functional degrees of freedom in the gauge field be encoded in a metric perturbation and/or axio-dilaton and, second, how are the remaining kinematic structures in (\ref{YM1}) and (\ref{GR1}) related? We take these in reverse order, since the answer to the second question will inform the first.

\subsubsection*{Kinematics}
The double copy form of the kinematic structure is revealed simply by rearranging terms in the gravity amplitude (\ref{GR1}):
\be\label{M1-GR}
     {i\mathcal{M}_{(1)} =\kappa} \int\!\hat{\ud}^4 k\, \hat{\delta}^4(p + q - k) \bigg[
    h^{\mu\nu} (k) (p - \tfrac12 k)_\mu(p - \tfrac12 k)_\nu +\tfrac14 R_{(1)}(k) + \tfrac12 k^2 \phi_{(1)}(k)
    \bigg] \;.
\ee
in which
\be
\label{R1}
    R_{(1)}(k) = k^2 {h^\sigma}_\sigma(k) - k_\mu k_\nu h^{\mu\nu}(k) \;,
\ee
is (the Fourier transform of) the linearised Ricci scalar, $R=\kappa R_{(1)}$. The first term in the square brackets of (\ref{M1-GR}) is explicitly what we want -- two copies of the kinematic terms in the gauge theory amplitude. What remains should therefore vanish, which constrains the dilaton to be a function of the metric variables according to 
\begin{align}
    \phi_{(1)}(k) &= -\frac{1}{2k^2} R_{(1)}(k)  \label{dilaton-eom-1} \\
    &= -\tfrac12 {h^\sigma}_\sigma(k) +\frac{k_\mu h^{\mu\nu}(k) k_\nu}{2k^2}  \;. \label{dilaton-eom-2}
\end{align}
To understand this, recall that for the double copy of perturbative scattering amplitudes in vacuum, graviton polarisations $\epsilon^{\mu\nu}$ emerge as the symmetric traceless product of gluon polarisations,   and the dilaton emerges through the trace. 
{Here the dilaton is again essentially a trace, as seen in the first term of (\ref{dilaton-eom-2}), and we will see immediately below that the metric $h^{\mu\nu}$ appears as the product of gauge fields.} The second term in (\ref{dilaton-eom-2}), going like $khk$ would not arise for scattering of on-shell states in vacuum, as then $k\epsilon k = 0$. Here it may be nonzero, though, since our background may be sourced.

The origin of (\ref{dilaton-eom-1}) is readily identified -- the equations of motion for the dilaton which follow from the \emph{gravitational} part of the original action are, assuming a vanishing axion,
\begin{equation}\label{dilaton-eom-full}
	{\sqrt{-g} e^{\phi} \bigg[- 2 R - 2 g^{\mu\nu} \pd_\mu\phi \pd_\nu\phi
 \bigg]
 + 4 \pd_\mu \big(\sqrt{-g} e^{\phi} g^{\mu\nu} \pd_\nu \phi  \big) = 0\;.}
\end{equation}
Linearising this equation and going to Fourier space, one immediately recovers (\ref{dilaton-eom-1}); hence the dilaton is constrained to obey the equations of motion in a prescribed metric and with no matter fields.
We similarly take the axion to obey its equation of motion, for which $\chi=0$ is always a solution, and we adopt this from here on.
The potential for cancellations between the dilaton and metric contributions is made \emph{explicit} by our choice of field variables in (\ref{DC-L-2}), and is a benefit of that choice. Had we used the original variables of (\ref{DC-L-1}) we would have had to allow the metric to depend on the dilaton, and then separate out a trace-like term to find the same cancellation. Our change of variables seems to do this for us.

\subsubsection*{Functional degrees of freedom}
We expect, from the form of double copy in flat space, that the product of gauge fields \emph{in Fourier space} should yield the Fourier transform of the metric perturbation, $h(k)\sim A(k)A(k)$. This is in the spirit of the {convolutional double copy~\cite{Anastasiou:2014qba,Anastasiou:2018rdx}} (i.e.~the metric should be a convolution of gauge fields in position space), which has been applied in Minkowski~\cite{Luna:2020adi} and in (A)dS~\cite{Liang:2023zxo}, see also~\cite{Borsten:2019prq,Borsten:2020xbt}.  (For special cases where the double copy is local in position space, suggesting a convolution in Fourier space, see~\cite{Luna:2022dxo}.)

Dimensional analysis requires, however, the presence of some other structure which cancels the dimensionality of one copy of the gauge field. This suggests the introduction of a biadjoint scalar (BAS) field~$\Phi$, {see e.g.~\cite{Cachazo:2013iea,Anastasiou:2014qba,Liang:2023zxo}, such that}
\begin{equation}\label{h=AA}
    h^{\mu\nu}(k) = A^{\mu a}(k) \Phi^{-1}_{ab}(k) A^{\nu b}(k).
\end{equation}
There is a great deal of flexibility in choosing $\Phi$, and one could also take the product of different gauge fields $A^\mu$ and ${\tilde A}^\mu$ as $h\sim A\Phi^{-1}{\tilde A}$, a freedom analogous to choosing two different gauge groups when taking the zeroth copy. Exploiting these freedoms may be useful for simplifying calculations but does not affect which metrics can be obtained through (\ref{h=AA}) -- any choice of $\Phi$ can ultimately be absorbed into a transformation of the gauge field.

With this relation between the metric perturbation and gauge field, we can obtain the correct gravity amplitude from (\ref{YM1}) by replacing $g$ with $\kappa/2$, familiar from the literature, and by replacing the colour factor with a copy of the  {remainder of the} amplitude,
\begin{equation}
    \sfT^a_{ij} \to - \Phi^{-1}_{ab}(k) A^{b \nu}(k) \big(p    - \tfrac12 k\big)_\nu \;,
\end{equation}
tied to the original amplitude by the BAS. Much like in vacuum the coupling, delta functions and momentum integrals are not copied --  {here and below, our double copy prescription will work at the level of \emph{integrands}.} 
Note that we have let $g \to \kappa/2$ \textit{before} copying the numerator; as such there is a factor of 2 in the amplitude that is effectively not copied, but this is equivalent to instead defining the colour factor in terms of generators rescaled by $\sqrt{2}$, as is typical~\cite{Bern:2019prr}, and copying all numerical factors.

We comment briefly on the sources of our gauge field and metric (\ref{h=AA}), since less is known about sourced solutions, compared to vacuum solutions, in classical double copy, see though~\cite{Easson:2021asd}. If the gauge background is sourced by $J_\mu$ we may write, at the linearised level, $A_\mu^a(k) = 
    \big(k_\mu k_\nu A^{a\nu}(k)-J_\mu^a(k) \big)/k^2$. Substituting into (\ref{h=AA}), and writing e.g.~$h=A\circ A$ to compactify notation for products with the BAS~\cite{Liang:2023zxo}, the Einstein equations for the metric (\ref{h=AA}) become 
\be
\begin{split}
    G_{\mu\nu}(k)
    &=
\frac{1}{2k^2}\bigg[    J_{\mu}(k) \circ { J}_{\nu}(k) - \bigg( \eta_{\mu\nu} - \frac{k_\mu k_\nu}{k^2}\bigg) J_\alpha(k) \circ { J}^\alpha(k) \bigg]\;. 
\end{split}    
\ee
The term in round brackets is the $k$-transverse propagator, which appears because $k\cdot J=0$ in the gauge theory, and guarantees that $k_\mu G^{\mu\nu} = 0$. (For discussion of how gravitational symmetries emerge from gauge theory via double copy see~\cite{Anastasiou:2014qba} and \cite{Ferrero:2024eva}.)

\subsubsection*{Summary}
Beginning with a background gauge field, we can form a background metric from the product of gauge fields in Fourier space, in the spirit of the convolutional double copy. The metric thus inherits all functional degrees of freedom originally present, and we have seen that the dilaton is then fixed via its equations of motion. Scattering massive fields on these gauge and gravitational backgrounds, we find that the leading order amplitudes indeed obey a double copy relation. The gravitational amplitude is, finally,
\be\label{M1-GR-final}
    {i\mathcal{M}_{(1)} =
    \kappa} \int\!\hat{\ud}^4 k\, \hat{\delta}^4(p + q - k) h^{\mu\nu}(k) \Big[(p - \tfrac12 k)_\mu(p - \tfrac12 k)_\nu
    \Big] \;,
\ee
which is to be compared against the gauge theory amplitude (\ref{YM1}). {Identifying $h$ as in (\ref{h=AA})}, the double copy structure is evident.

It is interesting to note how the choice of colour structure in $A_\mu$ affects which metrics can be obtained through (\ref{h=AA}). If the gauge field contains only a single generator, for example, the resulting metric perturbation is necessarily a bivector. Constructing the most general metric would seem to require as many generators as there are spacetime dimensions. There are then many different choices of $A_\mu$ which will yield the same metric.

The question to ask next is whether these double copy results persist beyond the linearised approximation and, if so, what new structures arise?  In particular we note that the convolutional double copy is only defined at the linearised level, which makes the question of how functional degrees of freedom double copy, beyond leading order, especially interesting.  These are the topics of the next section.
%

\section{Next to leading order}\label{sec:NLO}
%
Here we generalise the calculations above to next to leading order (NLO) in perturbation theory. As before we present the individual gauge and gravity calculations first, then discuss how to relate them via double copy.
\subsection{Gauge theory}
%
It is straightforward to compute the NLO gauge theory amplitude using the Feynman rules in Fig.~\ref{FIG:FEYN-YM}. Recall that momentum is conserved at each vertex, but that there is an accompanying integral over the background four-momentum. We continue to leave all such integrals unevaluated and therefore introduce for them the compact notation
\begin{equation}
    \int_n \equiv \int\! 
      \prod_{i = 1}^n \hat{\ud}^4 k_i
      \,\, \deltahat^4\bigg( p + q  - \sum_{j = 1}^n k_j \bigg) \;.
\end{equation}
This also allows us to focus on the \emph{integrands}. With this, the NLO gauge theory amplitude is easily found to be
\be\label{YM2}
	{i\mathcal{A}_{(2)}} = 
 {-(2g)^2}
 \int_2 \frac{{(\sfT^a \sfT^b)_{ij}} }{\rho} A^{\mu a}(k_1) A^{\nu b}(k_2) \left[  \left( p  - \tfrac{1}{2}k_1\right)_\mu\left(q - \tfrac{1}{2}k_2\right)_\nu     + \frac{\rho}{4} \eta_{\mu\nu} \right] \;,
\ee
where $\rho \equiv (p - k_1)^2 - m^2$ is the propagator denominator. The first term in (\ref{YM2}) comes from the single diagram with two three-point vertices -- as each interaction is with the background, there is no distinction between different orderings of gluon legs.
The second term comes from the seagull diagram. The amplitude (\ref{YM2}) exhibits a particularly simple colour structure, which generalises directly to higher orders, see~\cite{Johansson:2015oia} and \cite{Bjerrum-Bohr:2019nws} for more comments on this `single quark line' type colour factor. {Gauge invariance is readily checked. The $\mathcal{O}(g^2)$ amplitude, that is the sum $\mathcal{A}_{(1)}+\mathcal{A}_{(2)}$, is invariant up to terms of order $g^3$ under (\ref{gauge-x-form-full}); the nonlinear part of the transformation of (\ref{YM1}) cancels against the linearised transformation of (\ref{YM2}).}

Diagrams which may be neglected, or contribute irrelevant phases, in vacuum (e.g.~bubbles, tadpoles, disconnected contributions), can contain nontrivial physical information in backgrounds (on e.g.~vacuum decay and self focussing). Disconnected loop diagrams can contribute to our two-point amplitude already at order $\kappa^2$ but we expect that these pieces can be double-copied separately; here we will focus on the connected tree-level piece to pair production.

\subsection{Gravity}
%
The analogous NLO calculation in gravity is only complicated by the fact that, as we move beyond the linearised approximation, the amplitude will pick up contributions from higher order corrections to the metric (and dilaton). This requires a slightly more careful counting of corrections than in gauge theory.

From here on we work with a general perturbation to Minkowski spacetime, {again without choosing a particular gauge,} which we specify through the inverse metric
\begin{equation}
    g^{\mu\nu}(x) = \eta^{\mu\nu} - \h^{\mu\nu}(x),
\end{equation}
in which the perturbation may now contain (in preparation for both the NLO calculation and higher-order calculations below) terms of all orders $\kappa$,
\begin{equation}\label{all-orders-metric-general}
    \h^{\mu\nu}(x) \equiv \sum_{n= 1}^\infty \kappa^{n} h^{\mu\nu}_{(n)}(x) \;.
\end{equation}
(The perturbations $h^{\mu\nu}_{(n)}(x)$ can readily be related order by order to a general perturbation of the metric $g_{\mu\nu}$.)  Note that to avoid clutter we will when possible omit the subscript on the linear piece, writing $h_{(1)}^{\mu\nu}\equiv h^{\mu\nu}$, which also matches the notation of Sec.~\ref{sec:leading}. Given the relations (\ref{dilaton-eom-2}) and (\ref{dilaton-eom-full}) seen above, we similarly expand the dilaton as
\begin{equation}
    \phi(x) = \sum_{n=1}^\infty \kappa^n \phi_{(n)} (x) \;.
\end{equation}
We expand the Lagrangian in (\ref{quad-action}) up to second order in $\h$ and $\phi$, which is enough to capture all terms which can contribute to NLO:
\begin{equation}\label{L-order-2}
\begin{split}
    \mathcal{L} 
    =\,
    &\mathcal{L}_0 \\
    &+\Big(  \frac{1}{2}\h + \phi \Big)  \mathcal{L}_0
    - \h^{\mu\nu} \pd_\mu \varphi^\dagger \pd_\nu \varphi \\
    &\quad + 
    \bigg( \frac14 {\h^{\mu\nu} \h_{\mu\nu} } + \frac18 \h^2 + \frac12 {\h \phi} + \frac12 {\phi^2} \bigg) 
    \mathcal{L}_0
    - \Big(  \frac12 {\h} + \phi \Big) \h^{\mu\nu} \pd_\mu \varphi^\dagger \pd_\nu \varphi
     {\, +\ldots}
    \;,
\end{split}
\end{equation}
in which $\mathcal{L}_0$ is the free scalar Lagrangian in flat space, and indices are raised and lowered using the Minkowski metric as usual.
\begin{figure}[t!]
\begin{equation*}
\begin{split}
	\begin{tikzpicture}[baseline=(o.base)]
		\begin{feynman}[small]
			\vertex (o) at (0,0);
			\vertex[above right=of o] (s1);
			\vertex[below right=of o] (s2);
			\vertex[left=of o] (b){$\circ$};			
			\diagram* {
				(o) -- [fermion, momentum = $p$] (s1),
				(o) -- [anti fermion, momentum' = $q$] (s2),
				(b) -- [photon] (o),
			};
		\end{feynman}
	\end{tikzpicture} 
	&= i  \h^{\mu\nu}(k) \left[ p_\mu q_\nu   -  \frac12 \left(p \cdot q + m^2  \right)\eta_{\mu\nu}     \right]
 \\
	\begin{tikzpicture}[baseline=(o.base)]
		\begin{feynman}[small]
			\vertex (o) at (0,0);
			\vertex[above right=of o] (s1);
			\vertex[below right=of o] (s2);
			\vertex[left=of o] (b){$\circ$};			
			\diagram* {
				(o) -- [fermion, momentum = $p$] (s1),
				(o) -- [anti fermion, momentum' = $q$] (s2),
				(b) -- [scalar] (o),
			};
		\end{feynman}
	\end{tikzpicture}
	&= - i \phi(k) \left( p \cdot q + m^2  \right), 
\end{split} 	
\end{equation*}
\caption{\label{FIG:FEYN-GR2} Feynman rules for a complex scalar interacting with an all orders background graviton and dilaton.} 
\end{figure}
The Feynman rules generated by the \emph{second} line of (\ref{L-order-2}) are formally the same as those in Fig.~\ref{FIG:FEYN-GR1} except that the background $\h$ now contains all orders in $\kappa$, which we represent graphically with a circle `$\circ$' on the external leg in Fig.~\ref{FIG:FEYN-GR2}. So, this diagram evaluated on-shell, for example, gives contributions \textit{at every order} in~$\kappa$. 

The leading order Feynman rules of Fig.~\ref{FIG:FEYN-GR1} are recovered by extracting the linear piece $\kappa h^{\mu\nu}_{(1)}$ of the full perturbation, denoted by replacing the circle with a cross in the diagram. More generally we have e.g.~
\begin{equation}
    	\begin{tikzpicture}[baseline=(o.base)]
		\begin{feynman}[small]
			\vertex (o) at (0,0);
			\vertex[above right=of o] (s1);
			\vertex[below right=of o] (s2);
			\vertex[left=of o] (b){$\circ$};			
			\diagram* {
				(o) -- [fermion] (s1),
				(o) -- [anti fermion] (s2),
				(b) -- [photon] (o),
			};
		\end{feynman}
	\end{tikzpicture}
    =
    	\begin{tikzpicture}[baseline=(o.base)]
		\begin{feynman}[small]
			\vertex (o) at (0,0);
			\vertex[above right=of o] (s1);
			\vertex[below right=of o] (s2);
			\vertex[left=of o] (b){$\times$};			
			\diagram* {
				(o) -- [fermion] (s1),
				(o) -- [anti fermion] (s2),
				(b) -- [photon] (o),
			};
		\end{feynman}
	\end{tikzpicture}
    +
    	\begin{tikzpicture}[baseline=(o.base)]
		\begin{feynman}[small]
			\vertex (o) at (0,0);
			\vertex[above right=of o] (s1);
			\vertex[below right=of o] (s2);
			\vertex[left=of o] (b){$\times\times$};			
			\diagram* {
				(o) -- [fermion] (s1),
				(o) -- [anti fermion] (s2),
				(b) -- [photon] (o),
			};
		\end{feynman}
	\end{tikzpicture} 
    +
    	\begin{tikzpicture}[baseline=(o.base)]
		\begin{feynman}[small]
			\vertex (o) at (0,0);
			\vertex[above right=of o] (s1);
			\vertex[below right=of o] (s2);
			\vertex[left=of o] (b){$\times\!\times\!\times$};			
			\diagram* {
				(o) -- [fermion] (s1),
				(o) -- [anti fermion] (s2),
				(b) -- [photon] (o),
			};
		\end{feynman}
	\end{tikzpicture} 
    + \; \cdots,
\end{equation}
with each term on the right hand side picking out the contribution at a specific order in $\kappa$, {concretely}
\begin{equation}
    	\begin{tikzpicture}[baseline=(o.base)]
		\begin{feynman}[small]
			\vertex (o) at (0,0);
			\vertex[above right=of o] (s1);
			\vertex[below right=of o] (s2);
			\vertex[left=of o] (b){$\underbrace{\times \cdots \times}_{n}$};			
			\diagram* {
				(o) -- [fermion, momentum = $p$] (s1),
				(o) -- [anti fermion, momentum' = $q$] (s2),
				(b) -- [photon] (o),
			};
		\end{feynman}
	\end{tikzpicture} 
	= i \kappa^n h_{(n)}^{\mu\nu}(k) \bigg[ p_\mu q_\nu   -  \frac12 \left(p \cdot q + m^2  \right)\eta_{\mu\nu}     \bigg] \;.
\end{equation}
This notation allows us to visually keep track of the perturbation parameter through the number of crosses in a diagram. 
\begin{figure}[t!]
\begin{equation*}
\begin{split}
		\begin{tikzpicture}[baseline=(o.base)]
		\begin{feynman}[small]
			\vertex (o) at (0,0);
			\vertex[above right=of o] (s1);
			\vertex[below right=of o] (s2);
			\vertex[above left=of o] (b1){$\circ$};
			\vertex[below left=of o] (b2){$\circ$};			
			\diagram* {
				(o) -- [fermion, momentum = $p$] (s1),
				(o) -- [anti fermion, momentum' = $q$] (s2),
				(b1) -- [photon] (o),
				(b2) -- [photon] (o),
			};
		\end{feynman}
	\end{tikzpicture} 
	&= i  \h^{\mu\nu}(k_1) \h^{\alpha \beta}(k_2) \left[ \frac{\eta_{\alpha \beta}}{2} p_\mu q_\nu  - \left(  \frac{\eta_{\alpha\mu} \eta_{\beta\nu}}{4} + \frac{\eta_{\mu\nu} \eta_{\alpha\beta}}{8} \right) \left( p \cdot q + m^2  \right)  \right]\\
	\begin{tikzpicture}[baseline=(o.base)]
		\begin{feynman}[small]
			\vertex (o) at (0,0);
			\vertex[above right=of o] (s1);
			\vertex[below right=of o] (s2);
			\vertex[above left=of o] (b1){$\circ$};
			\vertex[below left=of o] (b2){$\circ$};			
			\diagram* {
				(o) -- [fermion, momentum = $p$] (s1),
				(o) -- [anti fermion, momentum' = $q$] (s2),
				(b1) -- [photon] (o),
				(b2) -- [scalar] (o),
			};
		\end{feynman}
	\end{tikzpicture} 
	&= i  \h^{\mu\nu}(k_1)  \phi(k_2) \left[ p_\mu q_\nu   - \frac{\eta_{\mu\nu}}{2}\left(p \cdot q + m^2  \right)     \right]\\
	\begin{tikzpicture}[baseline=(o.base)]
		\begin{feynman}[small]
			\vertex (o) at (0,0);
			\vertex[above right=of o] (s1);
			\vertex[below right=of o] (s2);
			\vertex[above left=of o] (b1){$\circ$};
			\vertex[below left=of o] (b2){$\circ$};			
			\diagram* {
				(o) -- [fermion, momentum = $p$] (s1),
				(o) -- [anti fermion, momentum' = $q$] (s2),
				(b1) -- [scalar] (o),
				(b2) -- [scalar] (o),
			};
		\end{feynman}
	\end{tikzpicture} 
	&= - \frac{i}{2} \phi(k_1) \phi (k_2) \left( p \cdot q + m^2  \right),
\end{split}
\end{equation*}
\caption{\label{FIG-FEYN-GR2-4point} Feynman rules that begin contributing to the pair creation amplitude at second order in the coupling. As before, replacing a circle on an external leg with $n$ crosses indicates taking the order $n$ contribution of the corresponding background (metric or dilaton).}
\end{figure}

At each perturbative order there is a new set of diagrams to take into account, built from a larger set of Feynman rules -- at NLO the four-point vertices in Fig.~\ref{FIG-FEYN-GR2-4point} are introduced, for example. They contribute along with the order $\kappa^2$ parts of the diagrams in Fig.~\ref{FIG:FEYN-GR2}. The remaining four tree-level diagrams are those with two three-point vertices.

Following the results found at leading order we take the dilaton to obey its equations of motion -- the leading piece $\phi_{(1)}$ is again replaced by the linearised Ricci scalar through (\ref{dilaton-eom-1}), and expanding (\ref{dilaton-eom-full}) to second order we have
\begin{equation}
\begin{split}
	4 \pd^2 \phi_{(2)} =  2 R_{(2)} + 2 \phi_{(1)} R_{(1)} &+ h R_{(1)} + 2   \pd \phi_{(1)} \cdot \pd \phi_{(1)}  \\
 &- 4 \pd_\mu \bigg[ \bigg(\frac{h}{2} \eta^{\mu\nu} +  \phi_{(1)} \eta^{\mu\nu} - h^{\mu\nu}  \bigg) \pd_\nu \phi_{(1)}    \bigg],
\end{split}
\end{equation}
which allows us to express $\phi_{(2)}$ in terms of the metric perturbations.

\subsection{Double copy}
%
Putting together all the diagrams and frequently utilising symmetries related to interchanges of Lorentz indices and momenta $k_i$ under the integral,
we find that the NLO amplitude is
\begin{equation}\label{GR2-proto}
\begin{split}
    & {i}\mathcal{M}_{(2)} =
     {\frac18}
    \kappa^2 \int_2 h^{\mu\nu} (k_1) h^{\alpha\beta} (k_2) \eta_{\nu\beta} \left( p - q \right)_\mu \left(p - q \right)_\alpha  
      {+\frac14}
     \kappa^2 \int_1 h_{(2)}^{\mu\nu}(k_1) \left( p - q \right)_\mu \left(p - q \right)_\nu \\
    & {+}
     \kappa^2\int_2 \frac{h^{\mu\nu}(k_1) h^{\alpha\beta}(k_2)}{\rho} \left[  \left( p  - \tfrac{1}{2}k_1 \right)_\mu   \left(q - \tfrac{1}{2}k_2  \right)_\alpha  +  \frac{\rho }{4} \eta_{\mu\alpha} \right]   \left[  \left( p  - \tfrac{1}{2}k_1 \right)_\nu   \left(q - \tfrac{1}{2}k_2  \right)_\beta     +   \frac{\rho }{4} \eta_{\nu\beta} \right].
\end{split}
\end{equation}
Compare this to (\ref{YM2}); the second line of the gravity amplitude displays a clear double copy structure.
Informed by the replacements at leading order we can obtain the {second line} from the gauge amplitude as follows:  leave the propagator, integral and factor of $(2g)^2$ untouched, and then replace the generators with a second copy of the remainder of the integrand,
\begin{equation}
    (\sfT^a \sfT^b)_{ij}  \to -  \Phi_{a a'}^{-1}(k_1) \Phi_{b b'}^{-1}(k_2)  A^{\alpha a'}(k_1)   A^{\beta b'}(k_2) \left[  \left( p  - \tfrac{1}{2}k_1\right)_\alpha\left(q - \tfrac{1}{2}k_2\right)_\beta     + \frac{\rho}{4} \eta_{\alpha\beta} \right],
\end{equation}
along with BAS fields. With the identification (\ref{h=AA}), the indices work out precisely so that we recover the second line of (\ref{GR2-proto}).

We are then left with the question of how to deal with the terms on line one of (\ref{GR2-proto}), in order to have a working double copy prescription. With our current setup the gravity amplitude matches our double copy prediction only if those terms vanish, which in turn requires
\begin{equation}\label{h2-constraint}
    h_{(2)}^{\mu\nu}(x) = - \tfrac12 h^{\mu\sigma}(x)\eta_{\sigma\rho} h^{\rho \nu}(x).
\end{equation}
Higher order corrections to the metric are thus \emph{determined} from the linear perturbation by enforcing a double copy structure. {As such, while the choice of gauge field and BAS still determines the metric, the story is not as straightforward as (\ref{h=AA}) suggests -- we will comment further on this result, and the resummation of the perturbative series, below.}

A notable special case is Kerr-Schild spacetimes~\cite{Monteiro:2014cda,Chawla:2023bsu,Easson:2023dbk}, where there are no higher order corrections to the metric and $h^{\mu\sigma}{h_{\sigma}}^{\nu}=0$ by construction, rendering (\ref{h2-constraint}) trivial.  Since in addition $h^\sigma_\sigma=0$ for Kerr-Schild, it seems likely that no further corrections are generated at higher orders in perturbation theory. {It would then follow that if $\kappa h$ is in} Kerr-Schild form, no resummation is necessary. {(This class includes plane waves, for which an all-orders double copy map has already found~\cite{Adamo:2017nia}.)}

We note that, instead of putting restrictions on the metric, the first line of (\ref{GR2-proto}) could be generated through the double copy if the gauge field and BAS have series expansions in~$g$. We will not pursue this option further here, though.

\section{Examples and all-order results}\label{sec:examples}

In this section we will use the above results to conjecture a double copy map at all orders. We will then identify a non-perturbative way to greatly simplify the dilaton contribution to our two-point amplitudes, which we will make use of when we turn to particular choices of background fields. This will allow us to explicitly extend the LO and NLO results above to higher orders in perturbation theory, and provide further all-orders conjectures.

\subsection{All-orders conjecture}

As is apparent from the Feynman rules of Fig.~\ref{FIG:FEYN-YM}, the tree-level amplitude at order $g^m$ of a single quark line interacting with a background can always be written as 
\begin{equation}\label{YM-order-n}
    i \mathcal{A}_{(m)} = g^m \int_m \frac{\prod_{s = 1}^m A^{\mu_s a_s}(k_s)}{D_m} N_{\mu_1 \dots \mu_m}  \Big(\textstyle{\prod}_{s=1}^m t^{a_s}\Big)_{ij} \;,
\end{equation}
where $D_m$ is the product of propagator denominators from the diagram with $m$ three-point vertices, $t^a = \sqrt{2} \sfT^a$ are generators normalised as in the vacuum double copy formalism, and $N_{\mu_1 \dots \mu_m}$ is some tensor built from kinematic variables. Following the procedure we have seen at leading and next to leading order, we make the replacements
\begin{equation}\label{replacement-order-n}
    g \to \frac{\kappa}{2} \;,
    \qquad
    {\Big({\textstyle{\prod}}_{s=1}^m \, t^{a_s}\Big)_{ij}}
 \to {\Big(\textstyle{\prod}_{s = 1}^m \Phi_{a_s b_s}^{-1}(k_s) A^{\nu_s b_s}(k_s) \Big) N_{\nu_1 \dots \nu_m}}\;,
\end{equation}
and, identifying the metric perturbation through (\ref{h=AA}), we arrive at a conjectured double copy amplitude
\begin{equation}
    i \mathcal{M}_{(m)} = \left( \frac{\kappa}{2} \right)^m \int_m \frac{\prod_{s=1}^m h^{\mu_s \nu_s}(k_s) }{D_m} N_{\mu_1 \dots \mu_m} N_{\nu_1 \dots \nu_m}\;.
\end{equation}
This double copy map can be put into a particularly familiar form by introducing the multi-indices $\mathbf{a} = (a_1, \dots, a_s)$ and $ \mathbf{b} = (b_1, \dots, b_s)$, through which we define colour factors, kinematic numerators, {and an object built from BAS fields as}
\begin{equation}\label{c-n-phi}
    c_{\mathbf{a}} \equiv  \Big({\textstyle{\prod}}_{s=1}^m \, t^{a_s}\Big)_{ij} \;, \qquad n_{\mathbf{a}} \equiv  \Big(\textstyle{\prod}_{s = 1}^m A^{\mu_s a_s}(k_s)\Big)   N_{\mu_1 \dots \mu_m} \;, \qquad \Phi_\mathbf{ab}^{-1} \equiv \prod_{s = 1}^m \Phi^{-1}_{a_s b_s} (k_s) \;.
\end{equation}
In terms of these quantities the gauge theory amplitude and its proposed double copy are compactly expressed as 
\begin{equation}\label{DC-An-compact}
    i \mathcal{A}_{(m)} = g^m \sum_\mathbf{a} \int_m \frac{n_\mathbf{a} c_{\mathbf{a}}}{D_m} \;,\qquad i \mathcal{M}_{(m)} = \left( \frac \kappa2 \right)^m \sum_\mathbf{a, b} \int_m \frac{n_\mathbf{a} \Phi^{-1}_\mathbf{ab} n_\mathbf{b} }{D_m} \;.
\end{equation}
In this form the amplitudes mirror the structure of double copy in vacuum, and our replacement rules (\ref{replacement-order-n}) {now effectively substitute} colour factors for kinematic numerators through $c_\mathbf{a} \to \Phi^{-1}_\mathbf{ab} n_\mathbf{b}$, along with some extra structure in the form of BAS fields. 

The sum runs over all colour structures $\mathbf{a}$, meaning every combination of colour indices $(a_1, \dots a_n)$ is double copied separately. Note that in general the colour factors $c_\mathbf{a}$ are not unique due to commutation relations between the generators. {Compare to the situation in vacuum, where the Jacobi relations similarly allow us to move terms between kinematic numerators without changing the amplitude, but where colour-kinematic duality can be invoked to make a choice that yields a sensible double copy amplitude~\cite{Bern:2008qj,Bern:2010ue,Bern:2010yg}. It is as yet unclear how to generalise colour-kinematic duality to general backgrounds~\cite{Adamo:2018mpq,Armstrong:2020woi,Sivaramakrishnan:2021srm,Cheung:2022pdk} and consequently there is some ambiguity in identifying the kinematic numerators $n_\mathbf{a}$ in (\ref{c-n-phi}). This can affect the double copy amplitude, as we will illustrate in Sec.~\ref{sec:higherorders}.}

On the gravity side, the expansion parameter $\kappa h$, which is generated by the choice of gauge and BAS fields, is seemingly related to the full metric perturbation $g-\eta$ in a non-trivial way. We will investigate this in the next section and see, in an analysis of examples, that it is still possible to choose the linear perturbation so as to produce a specific metric of interest.

\subsection{Simplifying the dilaton contribution}\label{sec:ricci}
Define the field
\be
    w(x) \equiv \exp \bigg(\frac{\phi(x)}{2}\bigg) \;.
\ee
In terms of $w(x)$ the dilaton equation of motion (\ref{dilaton-eom-full}) simplifies considerably; it becomes the \emph{linear} equation
\be\label{f-eom}
    \square w(x) - \frac14 R(x) w(x) =0 \;, \qquad \text{where}\qquad \square \equiv \frac{1}{\sqrt{-g}}\partial_\mu \sqrt{-g}g^{\mu\nu}\partial_\nu \;.
\ee
This is the equation of motion for a massless scalar \emph{non-minimally coupled} to the metric with $\xi=-1/4$. We will now show that the dilaton contribution to two-point amplitudes can be traded entirely for the same non-minimal coupling of the matter field $\varphi$ to the metric -- this offers a considerable simplification when extending the previous calculations to higher orders, as we will do below.

The pair production amplitude we consider may also be computed using the perturbiner approach~\cite{Arefeva:1974jv,Abbott:1983zw,Jevicki:1987ax,Rosly:1996vr,Rosly:1997ap,Mizera:2018jbh,Lee:2022aiu}, which states that tree-level two-point amplitudes for a field $\varphi$ on a background are computed by the quadratic part of the classical action evaluated on solutions of the equations of motion (chosen to implement the appropriate scattering boundary conditions). This result has recently been used in eikonal scattering~\cite{Adamo:2021rfq}, scattering on various backgrounds~\cite{Adamo:2017nia,Adamo:2021jxz,Kol:2021jjc,Adamo:2021rfq}, celestial holography~\cite{Gonzo:2022tjm}, pair creation~\cite{Ilderton:2023ifn}, and cosmology~\cite{Maldacena:2002vr, Maldacena:2011nz,Aoki:2024bpj}.

For us, the equation of motion following from the action (\ref{quad-action}) is just the wave equation
\be\label{matter-sol}
(\square +m^2)\, \varphi = 0 \;.
\ee
Evaluating (\ref{quad-action}) on any solution to (\ref{matter-sol}) reduces, via integration by parts, to a boundary term. If $\varphi_p$ represents the solution of (\ref{matter-sol}) corresponding to a free particle of momentum $p_\mu$ in the asymptotic future, then the boundary term to calculate  {for pair production} is
\be\label{randterm}
    \int_S\!\ud S^\mu\, e^{iq\cdot x}{\overset{\leftrightarrow}{\partial}}_\mu \varphi_p(x) \;,
\ee
which is just the standard Klein-Gordon inner product on an asymptotic surface $S$ with timelike normal. Another way arrive at this result is to observe that overlaps like (\ref{randterm}) simply compute the various Bogoliubov coefficients relating asymptotic particle wavefunctions for in/out states, in other words amplitudes, see e.g.~\cite{FradkinBook}.

If we define $\pi(x) = w(x) \varphi_p(x)$, and choose boundary conditions such that $w(x)$ and $\ud S^\mu \partial_\mu w(x)$ vanish on the boundary $S$, it follows trivially that (\ref{randterm}) may be computed replacing $\varphi_p$ with $\pi$. Since we take $w$ to obey its equation of motion (\ref{f-eom}), it is easily checked that the new field $\pi$ obeys the equations of motion of a massive scalar with non-minimal coupling $\xi=-1/4$;
\be\label{eom-pi}
    (\square + m^2)\, \pi(x) - \frac14 R(x) \pi(x) = 0 \;.
\ee
The important point is that we can compute the pair production amplitude simply by solving~(\ref{eom-pi}) and calculating~(\ref{randterm}) -- at no point do we have to \emph{explicitly} solve the dilaton equations of motion.
{To compute our amplitudes we can thus drop the dilaton entirely and work instead with the action}
\be\label{act-is-act}
 \int\!\ud^4x\, \sqrt{-g} \big(g^{\mu\nu} \partial_\mu \pi^\dagger \partial_\nu \pi - m^2 \pi^\dagger\pi+\frac14 R \pi^\dagger\pi\big)\;.
\ee
This makes calculations significantly simpler, as we will exploit below.

Interestingly, the numerical value of the non-minimal coupling here is the \emph{same} as arises in the worldline approach (to scalar fields interacting with gravity) as a finite counterterm which remains after cancellation of divergences associated with the path integral measure. See~\cite{Bastianelli:2006rx} for a discussion and history. It might be interesting to see if the worldline coupling could be reinterpreted in terms of the dilaton.

\subsection{Example: pair creation in FRW spacetimes
}

A natural gravitational analogue of particle creation in a time-dependent field is particle creation in a time-dependent metric~\cite{Parker:1968mv,Parker:1969au,Parker:1971pt,DeWitt:1975ys}. We focus here on the example of FRW spacetime, beginning by asking how this fits into our proposed double copy scheme at the linearised level. 

Taking the FRW line element in the standard form
\begin{equation}
	\ud\tau^2 = \ud t^2 - a^2(t) \ud \bm{x}^2,
\end{equation}
for $a(t)$ the scale factor, the inverse metric can be written 
\begin{equation}\label{f-def-1}
	g^{\mu\nu} = \eta^{\mu\nu} - 
 \bigg(\frac{1}{a^2(t)} - 1\bigg){X_3^{\mu\nu}},
\end{equation}
in which we have introduced
\begin{equation}\label{f-def-2}
{X_3^{\mu\nu}} \equiv e^\mu_i e^\nu_i,
\end{equation}
for $\{e_1^\mu,e_2^\mu,e_3^\mu\}$ an orthonormal basis of spacelike vectors.
Note that the tensor $X_3^{\mu\nu}$ is constant and obeys $X_3^{\mu\sigma}\eta_{\sigma\rho}X_3^{\rho\nu} = -X_3^{\mu\nu}$. In order to obtain this structure via the proposed double copy map (\ref{h=AA}) we consider a gauge field of the form
\begin{equation}
	A_a^\mu = \zeta^i_a (t) e_{i}^\mu \;,
\end{equation}
in which the functions $\zeta^i_a(t)$ are naturally assumed to be only $t$-dependent given the metric of interest. With this, we identify a metric perturbation
\begin{equation}
\begin{split}
	h^{\mu\nu} 
	&=  A_a^\mu \Phi_{a b}^{-1} A_{b}^\nu \\
	&=  \big( \zeta^i_a \Phi_{a b}^{-1} \zeta^j_{b} \big) e_i^\mu e_j^\nu \;.
\end{split}
\end{equation}
We now choose $\zeta$ and $\Phi$ such that $\zeta^i_a \Phi_{a b}^{-1} \zeta^j_{b} \propto \delta_{ij}$ in order to generate the required tensor structure $X_3^{\mu\nu}$. One interesting choice, given that most examples of classical double copy are essentially abelian, see though~\cite{Armstrong-Williams:2022apo}, is to consider a gauge group with {at least the same} number of colours as spatial dimensions, for example SU(2) in $3+1$ dimensions; one may then take $\zeta^i_a(t) = \zeta(t) \delta_{ia}$ for $\zeta$ a scalar. The gauge field is hence
\be\label{FRW-single-copy}
    A^\mu = \zeta(t) \sfT^i e_i^\mu \;,
\ee
and if we additionally choose $\Phi_{a b}^{-1} = \Phi^{-1} \delta_{a b} $ diagonal, then the metric perturbation is
\begin{equation}\label{h-frw-x}
	h^{\mu\nu} = \Phi^{-1} \zeta^2(t) X_3^{\mu\nu}\;,
\end{equation}
which has the desired tensor form.

{We now choose the scalar functions such that $\Phi^{-1} \zeta^2(t) = f(t)$; our general results then immediately relate the gauge and gravity amplitudes up to NLO, and (\ref{h2-constraint}) gives the FRW metric}
\be\label{eq:FRW-first}
    g^{\mu\nu} = \eta^{\mu\nu} - X_3^{\mu\nu}(\kappa f(t) + \frac12 \kappa^2 f^2(t) + \ldots) \;.
\ee
For a given scale factor $a(t)$, we would like to choose $\kappa f(t)$ such that the resummed metric matches (\ref{f-def-1}). This requires knowledge of how higher order corrections to the metric relate to the linear perturbation, which we investigate below.
\subsection*{Higher orders}\label{sec:higherorders}
Since we have restricted to a specific metric, we are able to push our amplitude calculations to the next order in perturbation theory. This illustrates some of the subtleties which can appear beyond NLO.

The $\mathcal{O}(g^3)$ contribution to the gauge theory amplitude in the background (\ref{FRW-single-copy}) is
\be\label{FRW-gauge-3}
   i\mathcal{A}_{(3)} = -(2g)^3 \sfT^i \sfT^j \sfT^k \int_3 \frac{\zeta(k_1) \zeta(k_2)\zeta(k_3)}{\rho_1 \rho_3} \bigg[ (e_i\cdot p)(e_j\cdot p)(e_k\cdot p) + {(e_i\cdot p)} \frac{\rho_3}{4} \delta_{jk}
   +{(e_k\cdot p)}\frac{\rho_1}{4}\delta_{ij}\bigg] \;,
\ee
as follows directly from the Feynman rules in Fig.~\ref{FIG:FEYN-YM}. Here $\rho_1 = (p-k_1)^2-m^2$ and $\rho_3 = (q-k_3)^2-m^2$ are the propagator denominators. The first term in (\ref{FRW-gauge-3}) comes from the diagram with three 3-point vertices, while the second and third terms comes from diagrams containing a four-point vertex. 

According to the {proposal (\ref{replacement-order-n})} the expected double copy amplitude is
\be\label{FRW-gravity-3}
    i \mathcal{M}_{(3)} = \kappa^3 \int_3 \frac{f(k_1) f(k_2)f(k_3)}{\rho_1 \rho_3} {\bf p}^2 \bigg[{\bf p}^4 + \frac12 (\rho_1 + \rho_3) {\bf p}^2 +\frac{3}{16} (\rho_1^2+\rho_3^2) + \frac18 \rho_1 \rho_3\bigg] \;,
\ee
in which we have used $(e_i\cdot p)(e_i \cdot p) = p_\mu X_3^{\mu\nu} p_\nu = {\bf p}^2$. 

We turn to the gravitational calculation. The only tensor structure which will be produced by the double copy map is $X_3^{\mu\nu}$, {we generalise (\ref{eq:FRW-first}) to}
\be\label{FRW-metric-ansatz}
    g^{\mu\nu} = \eta^{\mu\nu} - X_3^{\mu\nu} \sum\limits_{n=1}^\infty c_n \kappa^n f^n(t) \;,
\ee
with $c_1 = 1$, $c_2 = \tfrac12$ already fixed by (\ref{h2-constraint}), and the remaining constants $c_n$ to be determined. {The corresponding on-shell dilaton is easily found order-by-order, and is a function only of $t$. (We give the explicit form as part of a later example, see~(\ref{GTF-dilaton}).)}

A direct calculation in gravity recovers precisely the amplitude (\ref{FRW-gravity-3}) if the metric has
\be\label{FRW-order-3}
    c_3 = \frac{5}{16} \;.
\ee
So, as at second order, the metric perturbations are determined at third order in terms of the linearised perturbation, if the double copy is to hold. {We will, below, make an all-orders conjecture for the $c_n$.} First, though, we highlight a subtle point.

Observe that we could have used the fact that $\sfT^j \sfT^j$ is a quadratic Casimir to write (\ref{FRW-gauge-3}) as
\be\label{FRW-gauge-3-2}
     i \mathcal{A}_{(3)}  \overset{!}{=} -(2g)^3 \sfT^i \sfT^j \sfT^k \int_3 \frac{\zeta(k_1) \zeta(k_2)\zeta(k_3)}{\rho_1 \rho_3} \bigg[ (e_i\cdot p)(e_j\cdot p)(e_k\cdot p) + (e_i\cdot p)\frac{\rho_1+\rho_3}{4}\delta_{jk}\bigg] \;,
\ee
in which only the terms containing a 4-point vertex are affected. This would suggest that the double copied amplitude is
\be
    i \mathcal{M}_{(3)}  \overset{!}{=} \kappa^3 \int_3 \frac{f(k_1) f(k_2)f(k_3)}{\rho_1 \rho_3} {\bf p}^2 \bigg[{\bf p}^4 + \frac12 (\rho_1 + \rho_3) {\bf p}^2 +\frac{3}{16} (\rho_1+\rho_3)^2\bigg] \;,
\ee
which is \emph{not} the same as (\ref{FRW-gravity-3}). 
{This amplitude can still be recovered from an FRW metric, but now with a different coefficient $c_3 = 9/16$ in the third-order metric perturbation. In what follows, though}, we continue to use our initial approach, in which we do not use any group identities. One reason for this is that, as we are about to see, it will allow us to conjecture all-orders results.   

\subsection{Example: double copy of electric fields and FRW-like spacetimes}\label{sec:electric}
%
In this section we work with a family of three backgrounds $A_{\sfd}^\mu$, for $\sfd\in\{1,2,3\}$, where
\be
\label{GTF-single-copy}
    A_\sfd^\mu = \zeta(t) \sum_{i=1}^\sfd \sfT^i e_i^\mu \;.
\ee
{(Note that we continue to work in $3+1$ dimensions in each case.)} 
For $\sfd=3$ we recover the FRW case already considered. The case $\sfd=1$ is effectively abelian, since there is then only a single generator $\sfT^1$ in the game and $A_1^\mu$ represents a time-dependent (chromo) electric field with profile $\zeta(t)$ and polarisation $e_1^\mu$. (For formal equivalences between particle creation in electric fields and FRW spacetimes see e.g.~\cite{Rajeev:2019okd}.) The case $\sfd=2$ interpolates between the abelian and FRW cases. We again assume there are at least as many generators as spacetime dimensions.
Through (\ref{h=AA}) and (\ref{h2-constraint}) the double copy map will generate, for each $\sfd$, metric perturbations proportional to
\be
    X_\sfd^{\mu\nu} \equiv \sum_{i=1}^\sfd e^\mu_i e^\nu_i \;,
\ee
i.e.~the metric perturbation is diagonal {and nontrivial only in directions~$1$ to~$\sfd\leq 3$. $X_\sfd$ has the properties $X_\sfd^{\mu\sigma}\eta_{\sigma\rho}X_\sfd^{\rho\nu} = - X_\sfd^{\mu\nu}$ and $X_\sfd^{\mu\nu}\eta_{\mu\nu} = -\sfd$.}

Beginning with the gauge theory, the $\mathcal{O}(g^3)$ amplitudes are given by taking (\ref{FRW-gauge-3}) and restricting the sums over $i,j,k$ to run from $1$ to $\sfd$ only. {As in (\ref{FRW-gauge-3}) we take the result generated directly from the Feynman rules, without exploiting any properties of the gauge group.} It follows that the expected double copied gravitational amplitude is, at $\mathcal{O}(\kappa^3)$, 
\be\label{GFT-gravity-3}
    i \mathcal{M}_{(3)} = \kappa^3 \int_3 \frac{f(k_1) f(k_2)f(k_3)}{\rho_1 \rho_3} {\bf p}_\sfd^2 \bigg[{\bf p}_\sfd^4 + \frac12 (\rho_1 + \rho_3) {\bf p}_\sfd^2 +\frac{\sfd}{16} (\rho_1^2+\rho_3^2) + \frac18 \rho_1 \rho_3\bigg] \;,
\ee
in which
\be
    {\bf p}_\sfd^2 \equiv p\cdot X_\sfd \cdot p = \sum_{i=1}^\sfd p_i p_i \;.
\ee
We turn to the direct calculation of the $\mathcal{O}(\kappa^3)$ amplitude in the gravitational theory, with the metric ansatz
\be\label{GTF-metric-ansatz}
    g^{\mu\nu}_\sfd \equiv \eta^{\mu\nu} - X^{\mu\nu}_\sfd \sum\limits_{j=1}^\infty c_j \kappa^j f^j(t) \;,
\ee
where, as before, $c_1=1$, $c_2=1/2$ fixed by our general results, and the remaining $c_j$ to be determined. The corresponding dilaton solution to order $\kappa^3$ is
\be\label{GTF-dilaton}
 \phi(t) = 
    \frac{\sfd}{2}\kappa f(t)
    -\frac{\sfd}{8}  \int_{\infty}^t\!\ud s \int_{\infty}^s\!\ud u\, \kappa^2 f'(u)^2
    +\frac{\sfd}{12} (6 c_3-1)  \kappa^3 f(t)^3 + \ldots \;,
\ee
with boundary conditions chosen such that $\{\phi(t), \partial_t \phi(t)\} \to 0$ as $t\to \infty$. Evaluating the $\mathcal{O}(\kappa^3)$ gravitational amplitude via the Feynman rules, one recovers exactly (\ref{GFT-gravity-3}) for
\be\label{c3}
    c_3 = \frac{\sfd+2}{16} \;.
\ee
Now that we have several related examples, we are prompted to speculate on the behaviour of higher-order terms in the series (\ref{FRW-metric-ansatz}) (and the general series (\ref{all-orders-metric-general})). We expect that there will be a similar constraint to (\ref{h2-constraint}) at $\order{\kappa^3}$, i.e. that $h_{(3)} \sim h h h$, with some linear combination of terms having differently contracted indices on the right hand side. The explicit $d$-dependence in $c_3$ indicates that trace terms begin contributing to $c_j$ for $j>2$, whereas there is no trace term in the general result at $\mathcal{O}(\kappa^2)$. At higher orders, different powers of traces could give various dependencies on powers of $\sfd$, and it is not obvious what their coefficients are. However, motivated by the structure of the contact terms which appear in both the matter and axio-dilaton sectors of the original {Lagrangian} (\ref{DC-L-1}) we make the following conjecture for the resummed series (\ref{GTF-metric-ansatz}):

\be\label{eq:conjecture}
    g^{\mu\nu}_\sfd = \eta^{\mu\nu} - \frac{\kappa f(t)}{\Big(1-\frac{\sfd}{4}\kappa f(t)\Big)^{2/\sfd}} X_\sfd^{\mu\nu} \;.
\ee
It is easily checked that the expansion of (\ref{eq:conjecture}) recovers $c_1=1$, $c_2=1/2$ and (\ref{c3}). 

There are of course infinitely many functions which could reproduce the patterns in (\ref{GFT-gravity-3}), (\ref{GTF-dilaton}) and (\ref{c3}). To provide some small additional evidence for the conjecture (\ref{eq:conjecture}), we have checked that it continues to hold at the next order in perturbation theory, $\mathcal{O}(\kappa^4)$, for $\sfd=1$, i.e.~the double copy map holds provided the metric perturbation has $c_4 = 1/16$, which agrees with the fourth-order expansion of (\ref{eq:conjecture}). (For this we used the `Ricci form' of the action (\ref{act-is-act}) which simplifies calculations.)

Given this, let us comment on the structures in the resummed metric (\ref{eq:conjecture}). For $\sfd=2$ we have a resummed geometric series, while for the abelian case $\sfd=1$ the denominator is a square, which is when the form of the metric most strongly resembles that of the contact terms in the original Lagrangian (\ref{DC-L-1}). We comment that in the same case the kinematic parts of the amplitudes $\mathcal{M}_{(n)}$ also simplify to perfect squares, with (writing out only the relevant structures) $\mathcal{M}_{(1)} \sim {\bf p_1}^2$, $\mathcal{M}_{(2)} \sim ({\bf p}_1^2+\tfrac14 \rho_1)^2$ and, from (\ref{GFT-gravity-3}),  $\mathcal{M}_{(3)} \sim {\bf p}_1^2({\bf p}_1^2 + \frac14(\rho_1+\rho_3))^2$.

It would seem natural for well-behaved gauge backgrounds to map to well-behaved gravitational backgrounds, or perhaps that singularities should map to corresponding singularities, as for e.g.~the classical double copy of the Coulomb field to the Schwarzchild metric~\cite{Monteiro:2014cda}. This draws attention to the fact that, for all $\sfd$, the resummed perturbations (\ref{eq:conjecture}) appear to be singular for sufficiently large perturbations. However, observe that (\ref{eq:conjecture}) corresponds to a scale factor
\be\label{eq:invertible}
    \frac{1}{a^2(t)} = 1 + \frac{\kappa f(t)}{\Big(1-\frac{\sfd}{4}\kappa f(t)\Big)^{2/\sfd}} \;,
\ee
and it is easily verified that this relation is \emph{invertible} in the sense that there is a finite $\kappa f(t)$ for \emph{any} given finite $a^2(t)$. The apparent divergence in our inverse metric simply corresponds to taking the scale factor $a(t)\to 0$, which is obviously singular (while the zero on the right hand side of (\ref{eq:invertible}), i.e.~the zero of the inverse metric, corresponds to taking $a(t)\to \infty$). Thus, if (\ref{eq:conjecture}) holds, we are able to recover any FRW spacetime from the double copy map. This is reassuring since it suggests that the conditions, like (\ref{h2-constraint}), generated by the double copy map are not really constraints, but are rather delivering a metric in a particular parameterisation.  {Furthermore, our conjecture potentially offers an amplitudes-based definition of \emph{classical double copy} for FRW, a type O spacetime, whereas the majority of classical double copy results hold for Kerr-Schild or type D spacetimes~\cite{Luna:2018dpt}. It would therefore, in future work, be extremely interesting to further test, or verify, (\ref{eq:conjecture}).}



\section{Conclusions}\label{sec:conclusions}
We have considered non-trivial two-point amplitudes of massive particles on  {arbitrary} gauge and gravitational background fields. Working perturbatively in the coupling to the backgrounds, we have established a double copy map between the gauge and gravitational amplitudes. In this map the gauge background defines a linearised metric perturbation, from which higher order metric perturbations (and an associated dilaton) are constructed.

While working perturbatively with the lowest multiplicity amplitudes available is perhaps the simplest approach one could take to the question of  {if or how} double copy applies in  {general} backgrounds, it is nevertheless direct and exposes a surprising amount of structure, even at leading and next-to-leading order in perturbation theory.

Our results have given insight into classical and convolutional double copy. While the latter has only been formulated at the level of linearised perurbations, we have seen that there is a sense in which this is all that is needed -- demanding that the double copy holds order by order in perturbation theory constructs a metric from the linearised perturbation. Notably, we have for specific examples been able to conjecture all-orders results which generate e.g.~all FRW spacetimes through double copy. We also found a non-perturbative way to simplify the dilaton contribution to our amplitudes, replacing it by a non-minimal coupling. While we do not 
expect this replacement to hold at higher points or loops it is useful for the computation of two-point tree amplitudes on backgrounds, which are of broad interest~\cite{Witten:1998qj,Kol:2021jjc,Adamo:2021rfq,Adamo:2021jxz,Gonzo:2022tjm,Ilderton:2023ifn,Aoki:2024bpj}.

As we worked perturbatively, we made the assumption that all our backgrounds were asymptotically flat. This can be relaxed, though. For potentials/metric perturbations which are only gauge-equivalent to zero, we can deal with the resulting memory terms as in e.g.~\cite{Kibble:1965zza,Cristofoli:2022phh}. This may give some insight into the issues with static terms in the double copy, seen in~\cite{Plefka:2019hmz}. For backgrounds which are genuinely non-flat asymptotically, we may appeal to the perturbiner approach~\cite{Arefeva:1974jv,Abbott:1983zw,Jevicki:1987ax,Rosly:1996vr,Rosly:1997ap} in order to define analogues of scattering amplitudes (e.g.~boundary to boundary correlators in AdS~\cite{Witten:1998qj}).

There are many possible directions for future research, including extensions of our results to higher orders, loops, and other numbers of dimensions~\cite{Emond:2022uaf,CarrilloGonzalez:2022ggn},  {as well as other gauge theories}. The focus could initially be on specific backgrounds where calculations are simpler (the class of fields in~\cite{Tomaras:2000ag,Tomaras:2001vs,Ilderton:2023ifn} are one candidate) or which have features of interest, e.g.~horizons~\cite{Chawla:2023bsu,He:2023iew}. It would also be interesting to investigate the self-dual sector~\cite{ Monteiro:2011pc,Chacon:2020fmr,Campiglia:2021srh,Armstrong-Williams:2022apo,Monteiro:2022nqt,Borsten:2023paw,Brown:2023zxm}. Another obvious target is higher-point amplitudes of the matter and gauge fields, i.e.~scattering of particles \emph{with} radiation;
{the more complex colour structures of such amplitudes would allow us to start exploring colour-kinematic duality in non-flat backgrounds~\cite{Adamo:2018mpq,Armstrong:2020woi,Sivaramakrishnan:2021srm,Cheung:2022pdk}.}
{We hope that having general results, even to NLO, may help to better expose how gauge transformations and diffeomorphisms are related through double copy.} Finally, an open question is `what is the single copy of amplitudes in a \emph{given} gravitational background?' For the family of FRW-like spacetimes considered here, our results suggest that the answer is the gauge field (\ref{GTF-single-copy}), although, recall e.g.~(\ref{eq:invertible}), it is necessary to go to all orders to see this explicitly. 

We hope to explore some of these questions in future work.

\medskip

\noindent\textit{We thank T.~Adamo, D.~Akpinar, R.~Gonzo, S.~Klisch, C.~Schubert and C.~Shi for useful discussions. The authors are supported by the STFC consolidator grant ST/X000494/1, ``Particle Theory at the Higgs Centre'' (A.I.) and an STFC studentship (W.L.).}

\providecommand{\href}[2]{#2}\begingroup\raggedright\endgroup

\bibliographystyle{JHEP}

\end{document}